\documentclass[aoas,MSNbibl,nameyear,dvips]{arximspdf}
\usepackage{dcolumn}
\usepackage{accents}
\usepackage{graphicx}

\doi{10.1214/13-AOAS652} 
\volume{7}
\issue{3}
\pubyear{2013}
\firstpage{1562}
\lastpage{1592}

\makeatletter

\newcolumntype{d}[1]{D{.}{.}{#1}}

\newproclaim{definition}{Definition}[section]

\newcommand{\myspan}{\operatorname{span}}

\renewcommand{\underbar}{\underaccent{\bar}}

\makeatother

\begin{document}
\begin{frontmatter}

\title{Modeling and forecasting electricity spot prices: A~functional
data perspective}
\runtitle{Modeling and forecasting electricity spot prices}

\begin{aug}
\author[A]{\fnms{Dominik} \snm{Liebl}\corref{}\ead[label=e1]{liebl@statistics.uni-koeln.de}\ead[label=e2]{dliebl@uni-bonn.de}}
\runauthor{D. Liebl}
\affiliation{Universities of Cologne and Bonn}
\address[A]{Seminar f\"ur Wirtschafts\\
\quad und Sozialstatistik\\
Universit\"at zu K\"oln\\
Albertus-Magnus-Platz\\
50923 K\"oln\\
Germany\\
\printead{e1}\\
and\\
Statistische Abteilung\\
Universit\"at Bonn\\
Adenauerallee 24-26\\
53113 Bonn\\
Germany\\
\printead{e2}} 
\end{aug}

\received{\smonth{2} \syear{2012}}
\revised{\smonth{3} \syear{2013}}

%
\begin{abstract}
Classical time series models have serious difficulties in modeling and
forecasting the enormous fluctuations of electricity spot prices.
Markov regime switch models belong to the most often used models in the
electricity literature. These models try to capture the fluctuations of
electricity spot prices by using different regimes, each with its own
mean and covariance structure. Usually one regime is dedicated to
moderate prices and another is dedicated to high prices. However, these
models show poor performance and there is no theoretical justification
for this kind of classification. The merit order model, the most
important micro-economic pricing model for electricity spot prices,
however, suggests a continuum of mean levels with a functional
dependence on electricity demand.

We propose a new statistical perspective on modeling and forecasting
electricity spot prices that accounts for the merit order model. In a
first step, the functional relation between electricity spot prices and
electricity demand is modeled by daily price-demand functions. In a
second step, we parameterize the series of daily price-demand functions
using a functional factor model. The power of this new perspective is
demonstrated by a forecast study that compares our functional factor
model with two established classical time series models as well as two
alternative functional data models.
\end{abstract}

%

\end{frontmatter}

\section{Introduction}\label{i}
Time series of hourly electricity spot prices have peculiar properties.
They differ substantially from time series of equities and other
commodities because electricity still cannot be stored efficiently and,
therefore, electricity demand has an untempered effect on the
electricity spot price [\citet{Knittel2005}].

The development of models for electricity spot prices was triggered by
the liberalization of electricity markets in the early 1990s. Hourly
electricity spot prices are usually considered to be multivariate
($24$-dimensional) time series since for each day $t$ the $24$
intra-day spot prices are settled simultaneously the day before [\citet
{Huisman2007}].

However, classical time series models adopted for electricity spot
prices such as autoregressive, jump diffusion or Markov regime switch
models reduce the multivariate time series to univariate time series
either by taking daily averages of the 24 hourly spot prices [\citet
{weron2004modeling}, \citet{Kosater2006} and \citet{koopman2007}] or by
considering each hour $h$ separately [\citet{Karakatsani2008}]. These
unnatural aggregations and separations of the data necessarily come
with great losses in information.

Our model, a functional factor model (FFM), is not a mere adaption of a
classical time series model but is motivated by the data-generating
process of electricity spot prices itself. Pricing in power markets is
explained by the merit order model. This model assumes that the spot
prices at electricity exchanges are based on the marginal generation
costs of the last power plant that is required to cover the demand. The
resulting so-called merit order curve reflects the increasing
generation costs of the installed power plants. Often, nuclear and
lignite plants cover the minimal demand for electricity. Higher demand
is mostly served by hard coal and gas fired power plants.

Due to its importance, the merit order model is referred to as a
fundamental market model [\citet{Burger2008}, Chapter~4]. Essentially,
the consideration of this fundamental model yields to the superior
forecast performance of our FFM in comparison to state of the art time
series models and alternative functional data models.

It is important to emphasize that the merit order model is not a static
model. The merit order curve rather depends on the variations of the
daily prices for raw materials, the prices of CO$_2$ certificates, the
weather, plant outages and maintenance schedules of power plants.

The merit order curve is most important for the explanation of
electricity spot prices in the literature on energy economics and
justifies our view on the set of hourly electricity spot prices $\{
y_{t1},\ldots,y_{t24}\}$ of day $t$. We do not interpret them as
$24$-dimensional vectors but rather as noisy discretization points of a
smooth price-demand function $X_t$, which can be formalized as follows:
\[
y_{th}=X_t(u_{th})+\varepsilon_{th},
\]
where $u_{th}$ denotes electricity demand at hour $h$ of day $t$ and
$\varepsilon_{th}$ is assumed to be a white noise process.

\begin{figure}

\includegraphics{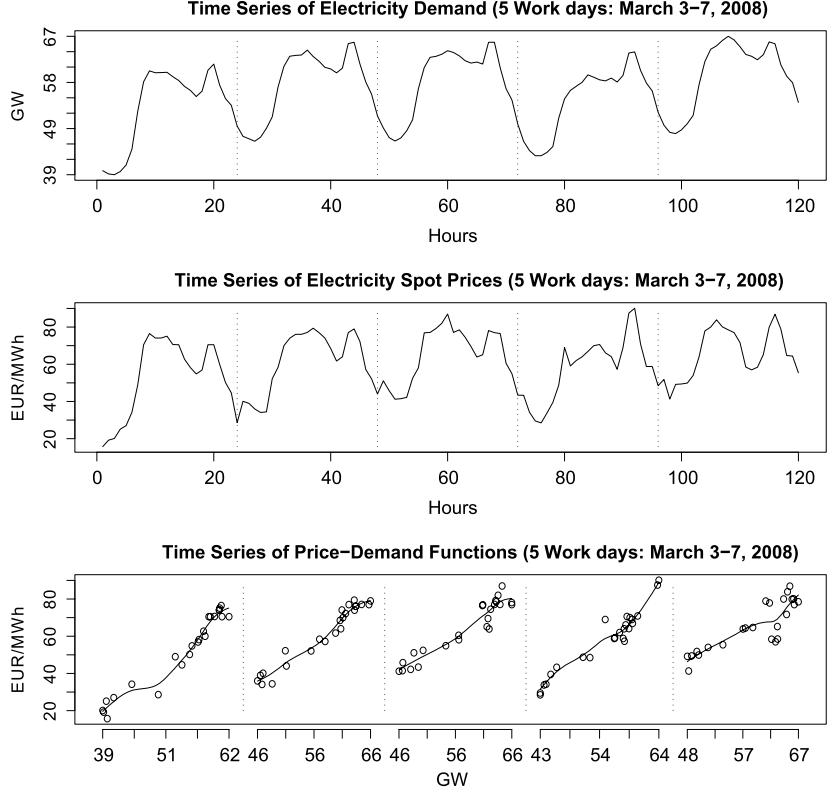}

\caption{\textsc{Upper panel}: time series of electricity demand
$(u_{th})$, measured in GW ($1$ GW $=1000$~MW). \textsc{Middle panel}:
electricity spot prices $(y_{th})$. \textsc{Lower panel}: price-demand
functions $(\hat{X}_t)$ with noisy discretization points
$(y_{t1},u_{t1}),\ldots,(y_{tN},u_{tN})$.} \label{Fig2}
\end{figure}

\begin{figure}

\includegraphics{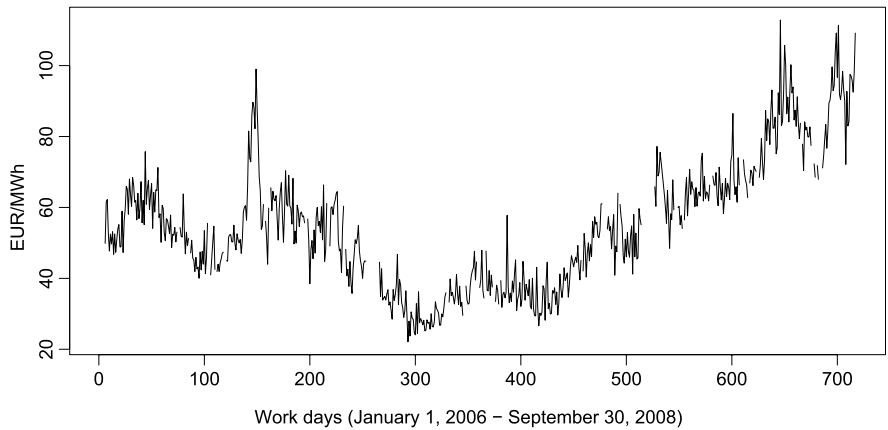}

\caption{Univariate time series of fitted price-demand functions $\hat
{X}_1(u),\ldots,\hat{X}_T(u)$ evaluated at $u=58\mbox{,}000$ MW. Gaps
correspond to holidays.}
\label{FigControlledPrices}
\end{figure}

The price-demand function $X_t(u)$ can be seen as the empirical
counterpart of the merit order curve estimated nonparametrically from
the $N=24$ hourly price-demand data pairs
$(y_{t1},u_{t1}),\ldots,(y_{tN},u_{tN})$. Five exemplary estimated
price-demand functions $\hat{X}_t(u)$ are shown in the lower panel of
Figure~\ref{Fig2}. Figure~\ref{FigControlledPrices} \mbox{visualizes} the
temporal evolution of the time series of price-demand functions by
showing the univariate time series $\hat{X}_1(u),\ldots,\hat{X}_T(u)$
for a fixed value of electricity-demand $u=58\mbox{,}000$ MW for the whole
observed time span of $T=717$ work days (Mo.--Fr.) from January 1, 2006
to September 30, 2008.

%
%

%
%

In order to capture the dynamic component of the price-demand
functions, we assume them to be generated by a functional factor model
defined as
\[
X_t(u)=\sum_{k=1}^K
\beta_{tk}f_k(u),
\]
where the factors or basis functions $f_k$ are time constant and the
corresponding scores $\beta_{tk}$ are allowed to be nonstationary time series.

We do not specify a constant mean function in our FFM, since we allow
the time series of price-demand functions $(X_t(u))$ to be
nonstationary. Consequently, the classical interpretation of the
factors $f_k$ as perturbations of the mean does not apply---as common
in the literature on dynamic (functional) factor models; see, for
example, \citet{hays2011functional}.

Note that the five price-demand functions in the lower panel of Figure~\ref{Fig2} are observed on different domains. This distinguishes our
functional data set from classical functional data sets, where all
functions are observed on a common domain. We refer to this feature as
\emph{random domains} and its consideration in Sections~\ref{rd} and
\ref{NoteonConvergence} is a central part of our estimation procedure.

We use a two-step estimation procedure. The first step is to estimate
the daily price-demand functions $\hat{X}_t$ by cubic spline smoothing
for all days $t\in\{1,\ldots,T\}$. The second step is to determine a
$K<\infty$ dimensional common functional basis system
$\{f_1,\ldots,f_K\}$ for the estimated price-demand functions
$\hat{X}_1,\ldots,\hat{X}_T$. Given this system of basis functions, we
model the estimated daily price-demand functions by a functional factor
model---using the basis functions as common factors. The fitted
discrete hourly electricity spot prices $\hat{y}_{th}$ are then
obtained through the evaluation of the modeled price-demand functions
at the corresponding hourly values of demand for electricity, formally
written as $\hat{y}_{th}=\hat{X}_t(u_{th})$.

Functional data analysis (FDA) can share our perspective on electricity
spot prices. A broad overview of many different FDA methods can be
found in the monographs of \citet{RamsayfdaBook2005} and \citet
{Ferraty2006}. Particularly, Chapter~8 in \citet{RamsayfdaBook2005} and
the nonparametric methods for computing the empirical covariance
function as proposed in \citet{Staniswal2010}, \citet{Yao2005}, \citet
{hall2006properties} and \citet{li2010uniform} are important
methodological references for this paper.

The application of models from the functional data literature to the
electricity market data is not new. For example, there is a vast
literature on modeling and forecasting electricity \emph{demand}; see,
for example, \citet{Ferraty2006} and \citet{Antoch}. However, modeling
and forecasting electricity spot prices is much more difficult than
modeling and forecasting electricity demand. The semi-functional
partial linear model (SFPL) of \citet{vilar2012forecasting} is one of
the very rare cases in which FDA methods are used to forecast
electricity spot prices.

Two very recent examples of other functional factor models are given by
the functional factor analysis in \citet{liufunctional} and the
functional dynamic factor model (FDFM) in \citet{hays2011functional}.
\citet{liufunctional} propose a new rotation scheme for the functional
basis components. \citet{hays2011functional} model a time series of
yield curves and estimate their model by the EM algorithm. In contrast
to the FDFM of \citet{hays2011functional}, we do not have to make a
priori assumptions on the stochastic properties of the time series of
scores in order to estimate our model components. Furthermore, we are
able to model and forecast functional time series observed on random domains.

Very close to the FDFM of \citet{hays2011functional} is the Dynamic
Semiparametric Factor Model (DSFM) of \citet{Park2009}. As our
functional factor model the DSFM does not need a priori assumptions on
the time series of scores. This and the fact that the DSFM was already
successfully applied to electricity prices [\citet{weron2008} and \citet
{Haerdle2010}] makes the DSFM a perfect competitor for our FFM.

The main difference between the FDFM of \citet{hays2011functional} and
the DSFM of \citet{Park2009} in comparison to our FFM is that the FFM
can deal with functional times series observed on random domains.
Furthermore, \citet{Park2009} use an iterating optimization algorithm
to estimate the basis functions of the DSFM, whereas we standardize the
elements of the time series $(X_t)$ so that we can robustly estimate
the basis functions by functional principal component analysis. Our
estimation procedure is much simpler to implement and faster with
respect to computational time than the Newton--Raphson algorithm
suggested in \citet{Park2009}.

The next section is devoted to the introduction of our data set and to
a critical consideration of the stylized facts of electricity spot
prices usually claimed in the electricity literature. In Section~\ref{FFM} we
present our functional factor model and in Section~\ref{e}
its estimation. An application of the model to real data is presented
in Section~\ref{a}. Finally, the performance of the functional factor
model is demonstrated by an extensive forecast study in Section~\ref{f}.

\section{Electricity data}\label{d}
We demonstrate our functional factor model by modeling and forecasting
electricity spot prices of the German power market traded at the
European Energy Exchange (EEX) in Leipzig. The German power market is
the biggest power market in Europe in terms of consumption. The
wholesale market is fragmented into an Over The Counter (OTC) market
and the EEX. While the OTC market has a continuous trade, the EEX has a
single uniform price auction with a gate closure for the day ahead
market at 12 p.m. the day before physical delivery. Although
three-fourths of the trading volume is settled via bilateral OTC
contracts, the EEX spot price is of fundamental importance as benchmark
and reference point for other markets, such as OTC or forward markets
[\citet{ockenfels2008electricity}, Chapter~1].

The data for this analysis stem from three different publicly available
sources. The hourly spot prices of the German electricity market are
provided by the European Energy Exchange (\href{http://www.eex.com}{www.eex.com}), hourly
values of Germany's gross electricity demand are provided by the
European Network of Transmission System Operators for Electricity
(\href{https://www.entsoe.eu}{www.entsoe.eu}), and German wind power infeed data are provided by the
EEX Transparency Platform (\href{http://www.transparency.eex.com}{www.transparency.eex.com}). The data
set used in our application is provided as part of the supplementary
material; see \citet{liebl2013}.

In the German electricity market, as in most of the electricity markets
in the world, renewable energy sources are usually provided with
purchase guarantees. Therefore, the hourly values of gross electricity
demand are not relevant for the pricing at the EEX but rather the
hourly values of gross demand minus the hourly electricity infeeds from
renewable energy sources. We consider only wind power infeed data since
the influences of other renewable energy sources such as photovoltaic
and biomass on electricity spot prices are still negligible for the
German electricity market (and their explicit consideration essentially
would lead to the same results).

The data consists of pairs $(y_{th},u_{th})$ with $y_{th}$ denoting the
electricity spot price and $u_{th}$ the electricity demand of hour
$h\in\{1,\ldots,24\}$ at day $t$. We define electricity demand
$u_{th}$ as the gross electricity demand of hour $h$ and day $t$ minus
the wind power infeed of electricity at the corresponding hour $h$ and
day $t$.

The data set analyzed in this article covers $T=717$ work days
(Mo.--Fr.) within the time horizon from January 1, 2006 to September 30,
2008. For the sake of clarity, only working days are considered in our
analysis since for weekends there are different compositions of the
power plant portfolio. The same reasoning applies to holidays and
so-called Br\"uckentage, which are extra days off that bridge single
working days between a bank holiday and the weekend. Therefore, we set
all holidays and Br\"uckentage to NA-values.

As a referee noted, the time span of our data set is peculiar. Starting
around January 2007, a price bubble for raw commodities such as coal
and gas was formed, which induced a strong increase in the electricity
spot prices. Interestingly, the increase in the electricity spot prices
is hardly visible in the original time series as shown in Figure~\ref{FigSpotPrices}. But it catches the eye in the plot of Figure~\ref{FigControlledPrices}, which shows the time series of price-demand
functions $(\hat{X}_t(u)|u)$ evaluated for a certain value of
electricity demand $u=58\mbox{,}000$ MW. The reason is that at this relatively
high value of electricity demand usually coal and gas fired power
plants cover the demand.

Very few (only $0.5\%$) of the data pairs $(y_{th},u_{th})$ with prices
$y_{th}>200$ EUR/MWh have to be treated as outliers since they cannot
be explained by the merit order model. Even in exceptional situations
the marginal costs of electricity production do not exceed the value of
$200$ EUR/MWh. Prices above this threshold are referred to as price
spikes and have to be explained using an additional scarcity premium
[\citet{Burger2008}, Chapter~4]. The analysis of price spikes is a
research topic on its own [\citet{Christensen2009}] and is not within
the scope of this paper.

We exclude the outliers for the estimation of our model and denote the
amount of data pairs of day $t$ used for estimation by $N_t\leq N=24$.
Nevertheless, we use the whole data set, including the outliers, in
order to assess the forecast performance of our model in Section~\ref{f}.

\subsection*{Review: Stylized facts of electricity data} Our functional
perspective on electricity spot prices allows us to review critically
the so-called ``stylized facts'' of hourly electricity spot prices
$(y_{th})$. Usually, time series of electricity spot prices are assumed
(i) to have deterministic daily, weekly and yearly seasonal patterns,
(ii) to show price dependent volatilities, and (iii) to be stationary
(after controlling for the seasonal patterns); see \citet
{HuismanRAndDeJong2003}, \citet{Knittel2005},
\citet{Kosater2006},
\citet{Huisman2007} and many others.

At first glance these stylized facts seem to be reasonable; see the
middle panel in Figure~\ref{Fig2}. However, the first two stylized
facts, (i) and (ii), are misleading since both have their origins in
the time series of electricity demand: the characteristics of
electricity demand are rather carried over to the time series of
electricity spot prices.

This can be explained by a micro-economic point of view, again using
the merit order model. The merit order curve induces a monotone
increasing supply function for electricity, which implies higher
electricity spot prices for higher values of electricity demand, where
electricity demand can be considered as inelastic. Given this
micro-economic point of view, we can regard the daily supply functions
for electricity as diffusers in the transmission from electricity
demand $u_{th}$ to the electricity spot price $y_{th}$.

Additional diffusion comes from the variations of the daily supply
functions caused by the varying input-costs of, for example, coal and
gas. Compare to this the time series of electricity demand with the
time series of electricity spot prices shown in the upper and middle
panels of Figure~\ref{Fig2}, respectively. The seasonal patterns of
electricity spot prices are just a diffused version of the smoother
seasonal patterns of electricity demand.

Price dependent volatility (ii) can be explained by the slope of the
merit order curve, which is increasing with electricity demand. Changes
in electricity demand have greater price effects for greater values of
electricity demand and therefore cause greater volatilities than is the
case for lower values of electricity demand.

Stationarity (iii) has to be considered critically, too. Recently,
\citet{Bosco2010} were able to show empirically that electricity spot
prices at the EEX have a unit root. The authors point out that the
stationarity assumption might be wrong in markets that are influenced
by price-enhancing sources such as prices for coal and gas since time
series of coal and gas prices are commonly found to be nonstationary.
Our functional factor model allows for nonstationarity in the time
series of price-demand functions $(X_t)$ and, in fact, tests indicate
that the estimated series of price-demand functions is nonstationary;
see Section~\ref{v}.

This short review of electricity spot prices demonstrates that
electricity data are complex with dynamics induced by the variations of
the merit order curve (mainly caused by varying input-costs) and
separate additional dynamics induced by electricity demand. To the best
of our knowledge, our functional factor model is the first model that
allows for a separate consideration of these two stochastic sources.
The variations dedicated to the dynamics of the merit order curve are
captured by the price-demand functions and modeled by our functional
factor model. The problem of modeling and forecasting electricity
demand is ``out-sourced'' and the statistician can choose powerful
specialized models for time series of gross electricity demand [\citet
{Antoch}] and time series of wind power [\citet{Lau2010}]. This
separation corresponds to the real data generating process.

\section{Functional factor model}\label{FFM}
As mentioned above, electricity spot prices $y_{t1},\ldots,y_{t24}$ are
actually one-day-ahead future prices since they are settled
simultaneously at day $t-1$. This implies that there is some degree of
uncertainty about the next day world in the electricity spot price
$y_{th}$, which we model nonparametrically as
%
%
\begin{equation}
\label{fm0} y_{th}=X_t(u_{th})+
\varepsilon_{th}.
\end{equation}
The error terms $\varepsilon_{th}$ are assumed to be i.i.d. white noise
errors with finite variance $\mathrm{V}(\varepsilon_{th})=\sigma
_\varepsilon^2$ and each function $X_t$ is assumed to be continuous
and square integrable.

For each function $X_t$ the values of electricity demand $u_{th}$ are
only observed within random sub-domains $\mathcal{D}(X_t)=[a_t,b_t]$,
where $[a_t,b_t]\subseteq[A,B]\subset\mathbb{R}$. The unobserved
univariate time series $(a_t)$ and $(b_t)$ are assumed to be time
series processes with $A\leq a_t<b_t\leq B$ and marginal p.d.f.s of $a_t$
and $b_t$ given by $f_a(z_a)>0$ and $f_b(z_b)>0$ for all $z_a,z_b\in
[A,B]$ and $t\in\{1,\ldots,T\}$.

The price-demand functions are relatively homogeneous. All of them look
very similar to the five randomly chosen price-demand functions shown
in the lower panel of Figure~\ref{Fig2}. The underlying reason for
this homogeneity is that, on the one hand, the merit order curve
induces rather simple monotone increasing price-demand functions. On
the other hand, the general portfolio of power plants, which is
reflected by the merit order curve, is changing very slowly and can be
considered as constant over the period of our analysis. We formalize
this homogeneity of the price-demand functions by the assumption that
the time series of price-demand functions $(X_t)$ is generated by a
functional factor model with time constant basis functions.

Given this assumption, every price-demand function $X_t$ can be modeled
by the same set of $K<\infty$ (unobserved) basis functions
$f_1,\ldots,f_k,\ldots,f_K$ with $f_k\in L^2[A,B]$, which span the
$K$-dimensional functional space $\mathcal{H}_K\subset L^2[A,B]$ such
that we can write
%
%
\begin{equation}\label{fm}
X_t(u)=\sum_{k=1}^K
\beta_{tk}f_k(u) \qquad\mbox{for all } u\in[a_t,b_t],
\end{equation}
where the common basis functions $f_k$ as well as the scores $\beta
_{tk}$ are unobserved and have to be determined from the data. We use
the usual orthonormal identification restrictions for the basis
functions, which require that $\int_A^Bf_k^2(u) \,du=1$ and $\int
_A^Bf_k(u)f_l(u) \,du=0$ for all $k<l\in\{1,\ldots,K\}$.

The $K$ real time series $(\beta_{t1}),\ldots,(\beta_{tK})$ are
defined as
%
%
\begin{equation}
\label{OLS} \pmatrix{\beta_{t1}
\cr
\vdots
\cr
\beta_{tK}}=
\pmatrix{\displaystyle \int_{a_t}^{b_t}f_1^2&
\cdots& \displaystyle \int_{a_t}^{b_t}f_1f_K
\vspace*{2pt}\cr
\vdots&\ddots&\vdots
\vspace*{2pt}\cr
\displaystyle \int_{a_t}^{b_t}f_1f_2&
\cdots& \displaystyle \int_{a_t}^{b_t}f_K^2}^{-1}
\pmatrix{\displaystyle \int_{a_t}^{b_t}f_1X_t
\vspace*{2pt}\cr
\vdots
\vspace*{2pt}\cr
\displaystyle \int_{a_t}^{b_t}f_KX_t}
\end{equation}
and are allowed to be arbitrary nonstationary processes. Note that for
$a_t=A$ and $b_t=B$ the definition of the scores $\beta_{tk}$
corresponds to the classical definition, given by $\beta_{tk}=\int
_{A}^{B}X_t(u)f_k(u) \,du$.

In the following section we propose an estimation algorithm for the
functional factor model.

\section{Estimation procedure}\label{e}
As outlined in Sections~\ref{i} and \ref{d}, we do not observe the
series $(X_t)$ directly but have to estimate each price-demand function
$X_t$ from the corresponding data pairs
$(y_{t1},u_{t1}),\ldots,(y_{tN_t},u_{tN_t})$. After this initial
estimation step, which is discussed in Section~\ref{e1}, we show in
Section~\ref{e2} how to determine an orthonormal $K$-dimensional basis
system $\{f_1,\ldots,f_K\}$ for the classical functional data case when
all price-demand functions $X_1,\ldots,X_T$ are observed on the
deterministic domain $\mathcal{D}(X_t)=[A,B]$. In Section~\ref{rd} we
generalize the determination of the orthonormal $K$-dimensional basis
system $\{ f_1,\ldots,f_K\}$ to our case, where the price-demand
functions\vspace*{1pt} $X_t$ are observed only on random domains
$\mathcal{D}(X_t)=[a_t,b_t]$. Finally, we define our estimator
$\{\hat{f}_1,\ldots,\hat{f}_K\}$ in Section~\ref{defestim}.

As usual for (functional) factor models, the set of factors $\{
f_1,\ldots,f_K\}$ in (\ref{fm}) is only determined up to
orthonormal rotations. Furthermore, the determination of an orthonormal
$K$-dimensional basis system $\{\hat{f}_1,\ldots,\hat{f}_K\}$ for a
given series $(\hat{X}_t)$ is, in the first instance, a mere algebraic
problem. But it is also a statistical estimation problem in the sense
that $\hat{\mathcal{H}}_K$, with $\hat{\mathcal{H}}_K=\myspan(\hat
{f}_1,\ldots,\hat{f}_K)$, is a consistent estimator of the theoretical
counterpart $\mathcal{H}_K$. The crucial assumption is that $X_t$
comes from the FFM (\ref{fm}). Consistency of the estimation follows
from the consistency of the single nonparametric estimators $\hat
{X}_t(u)$, which converge in probability against $X_t(u)$ as
$N_t\rightarrow\infty$ for all $u\in[a_t,b_t]$ and all $t\in\{
1,\ldots,T\}$ [\citet{benedetti1977nonparametric}]. Below in Section~\ref{NoteonConvergence} we consider this issue in more detail.

\subsection{Estimation of the price-demand functions $X_t$}\label{e1}
The estimation of the functions $X_t$ from the data pairs
$(y_{t1},u_{t1}),\ldots,(y_{tN_t},u_{tN_t})$ is done by minimizing
%
%
\begin{equation}
\label{spline}
d(\mathcal{X}|t)=\sum_{h=1}^{N_t}
\bigl(y_{th}-\mathcal{X}(u_{th}) \bigr)^2+b \int
_{a_t}^{b_t}\bigl(D^2\mathcal{X}(u)
\bigr)^2 \,du
\end{equation}
over all twice continuously differentiable functions $\mathcal{X}$,
where $D^2\mathcal{X}$ denotes the $2$nd derivative of $\mathcal{X}$
and $b>0$ is a preselected smoothing parameter. Spline theory assures
that any solution $\hat{X}_t$ of the minimization problem (\ref
{spline}) can be expanded by a natural spline basis [\citet
{deboor2001pgs}]. Therefore, we can use the expansion $\mathcal
{X}(u)=\mathbf{c}'\bolds{\phi}(t) $, where $\bolds{\phi}$
is the $(N_t+2)$-vector of natural spline basis functions of degree $3$
and $\mathbf{c}$ is the $(N_t+2)$-vector of coefficients over which
equation (\ref{spline}) is minimized. This procedure is usually
denoted as cubic spline smoothing and the interested reader is referred
to the monographs of \citet{deboor2001pgs} and \citet{RamsayfdaBook2005}.

An important issue that remains to be discussed is the selection of the
smoothing parameter $b$. Usually, the optimal smoothing parameter
$b^{\mathrm{opt}}$ is chosen by (generalized) cross-validation such that the
trade-off between bias and variance of the estimate $\hat{X}_t$ is
optimized asymptotically with respect to the mean\vspace*{1pt} integrated squared
error (MISE) criterion. However, our aim is not an optimal single
estimate $\hat{X}_t$ but rather an optimal estimation of the basis
system $\{f_1,\ldots,f_K\}$ for which we can use the information of all
price-demand functions $X_1,\ldots,X_T$.

Consequently, we do not have to optimize the MISEs of the single
estimators $\hat{X}_t$ but those of their weighted averages $\hat
{f}_1,\ldots,\hat{f}_K$. In this case an undersmoothing parameter
$\underbar{b}_K<b^{\mathrm{opt}}$ has to be chosen. This was discussed for the
first time in \citet{Benko2009a}. The underlying reason is that the
estimators $\hat{f}_1,\ldots,\hat{f}_K$ essentially are weighted
averages over all $\hat{X}_1,\ldots,\hat{X}_T$. Averaging reduces the
overall variance and therefore opens the possibility for a further
reduction in the MISEs of the estimators $\hat{f}_1,\ldots,\hat{f}_K$
by a further reduction of the bias-component through choosing
$\underbar{b}_K<b^{\mathrm{opt}}$ in the minimization of (\ref
{spline}). \citet{Benko2009a} propose to approximate an optimal
undersmoothing parameter $\underbar{b}_K$ by minimizing the following
cross-validation criterion:
%
%
\begin{equation}
\label{CVEq} \operatorname{CV}(b_K)=\sum
_{t=1}^T\sum_{i=1}^{N_t}
\Biggl\{y_{th}-\sum_{k=1}^K\hat
\gamma_{tk}\hat{f}_{k,-t}(u_{th}) \Biggr\},
\end{equation}
over $0\leq b_K\leq\infty$, where $\hat\gamma_{tk}$ are the OLS
estimators of $\hat\beta_{tk}$ and $\hat{f}_{k,-t}$ denote the
estimators of $f_{tk}$ based on the data pairs $(y_{sh},u_{sh})$ with
$s\in\{1,\ldots,t-1,t+1,\ldots,T\}$. We denote the estimators of $X_t$
based on an undersmoothing parameter $\underbar{b}_K$ by $\tilde
{X}_1,\ldots,\tilde{X}_T$ and those based on $b^{\mathrm{opt}}$ by $\hat
{X}_1,\ldots,\hat{X}_T$.

As can be seen in (\ref{CVEq}), an optimal undersmoothing
parameter $\underbar{b}_K$ depends on the dimension $K$. The problem
of choosing $K$ can be seen as a model selection problem, which
generally can be solved using information criteria. For our application
in Section~\ref{a} we use the simple cumulative variance criterion as
well as the AIC type criterion proposed in \citet{Yao2005}.

\subsection{Estimation of the basis system $\{f_1,\ldots,f_T\}$}\label{e2}
Our estimation procedure uses the property that any orthonormal basis
system $\{f_1,\ldots,f_K\}$ of the series $(X_t)$ has to fulfill the
minimization problem
%
%
\begin{equation}
\label{opt1} \sum_{t=1}^T\Biggl\|X_{t}-
\sum_{k=1}^K\beta_{tk}f_k\Biggr\|_2^2=
\min_{B_K}\sum_{t=1}^T
\min_{\gamma_{t1},\ldots,\gamma_{tK}\in\mathbb{R}}\Biggl\|X_{t}-\sum_{k=1}^K
\gamma_{tk}g_k\Biggr\|_2^2
\end{equation}
over all possible $K$-dimensional orthonormal basis systems $B_K=\{
g_1,\ldots,g_K\}$, where $g_1,\ldots,g_K\in L^2[A,B]$ and $\|\cdot\|_2$
denotes the functional L2 norm $\|x\|_2=\sqrt{\int_A^Bx^2(u) \,du}$ for
any $x\in L^2[A,B]$. This\vspace*{1pt} property is a direct consequence of the FFM
(\ref{fm}).

The minimization problem (\ref{opt1}) can be used to define an
estimator for a basis system $\{f_1,\ldots,f_K\}$. We would only have
to replace the unobserved functions $X_t$ with their undersmoothed
estimators $\tilde{X}_t$ and try to find a basis system that minimizes
the right-hand side (rhs) of (\ref{opt1}) using functional
principal component analysis (FPCA).

Before we present the analytic solution we adjust the minimization
problem~(\ref{opt1}). This adjustment yields a robustification, which
is needed since we allow the $K$ time series
$(\beta_{t1}),\ldots,(\beta_{tK})$ to be nonstationary processes. The
nonstationarity of the scores $(\beta_{t1}),\ldots,(\beta_{tK})$ implies
that different functions $X_t$ and $X_s$ can be of very different
orders of magnitude, that is, $\|X_t\|_2\ll\|X_s\|_2$. In such cases,
the squared L2-norm on the rhs of (\ref{opt1}) sets an overproportional
weight on functions with great orders of magnitude, and a functional
principal component estimator based on (\ref{opt1}) would be distorted
toward those functions $X_s$ that have great orders of magnitude
$\|X_s\|_2$.

If we were only interested in the determination of some set of basis
functions $\{f_1,\ldots,f_K\}$ that spans the same space $\mathcal
{H}_K$ as the set of functions $\{X_1,\ldots,X_T\}$, we would not have
to care about functions $X_s$ with great orders of magnitude $\|X_s\|$.
However, if we are interested in the interpretation of the basis
functions $f_k$, we want them to be representative for all functions
$X_1,\ldots,X_T$.

A general solution to this problem is to replace the price-demand
functions $X_t$ in (\ref{opt1}) with their standardized counterparts
$X_t^\ast=X_t/\|X_t\|$, which have equal orders of magnitude
$\|X_t^\ast\|=1$ for all $t\in\{1,\ldots,T\}$. Using this replacement
yields the following new minimization problem:
%
%
\begin{equation}
\label{opt2} \sum_{t=1}^T\Biggl\|X^\ast_{t}-
\sum_{k=1}^K\beta^\ast_{tk}f^\ast_k\Biggr\|_2^2
=\min_{B_K}\sum_{t=1}^T
\min_{\gamma_{t1},\ldots,\gamma_{tK}\in\mathbb{R}}\Biggl\|X^\ast
_{t}-\sum
_{k=1}^K\gamma_{tk}g_k\Biggr\|_2^2.
\end{equation}

Solving equation (\ref{opt2}) by FPCA generally will yield different
basis functions $f^\ast_k$ than solving (\ref{opt1}).
However, both minimization problems (\ref{opt1}) and (\ref{opt2}) are
equivalent in the sense that both sets of basis functions $\{f^\ast
_1,\ldots,f_K^\ast\}$ and $\{f_1,\ldots,f_K\}$ are equivalent up to
orthonormal rotations and therefore span the same space $\mathcal
{H}_K$. The standardization yields to a simple base change, which can
be seen by the fact that the original price-demand functions $X_t$ can
be written in terms of the basis functions $f^\ast_{k}$ as $X_t=\sum
_{k=1}^K(\|X_t\|\cdot\beta_{tk}^\ast) f^\ast_{k}$.

The standardization of all price-demand functions $X_t$ in the
minimization problem (\ref{opt2}) allows us to establish a
nondistorted estimator $\{\hat{f}_1,\ldots,\hat{f}_K\}$ that
represents all price-demand functions equally well. This approach is
similar to robust estimation procedures proposed by \citet
{Locantore1999} and \citet{Gervini2008} but differs conceptually
insofar as we do not consider any functional observation $X_t$ as an outlier.

We construct our estimator $\{\hat{f}_1,\ldots,\hat{f}_K\}$ from the
analytic solution of the minimization problem (\ref{opt2}). The
solutions of the inner minimization problem with respect to the scores
$\gamma_{tk}$ are given by least squares theory, and we can write
%
%
\begin{equation}
\label{opt3} \sum_{t=1}^T\Biggl\|X^\ast_{t}-
\sum_{k=1}^K\beta^\ast_{tk}f^\ast_k\Biggr\|_2^2=
\min_{B_K}\sum_{t=1}^T\bigl\|X^\ast_{t}-P^g_KX^\ast_t\bigr\|_2^2,
\end{equation}
where $P^g_K$, defined as $P^g_KX_t^\ast=\sum_{k=1}^K (\int
_{A}^{B}X_t^\ast(v)g_k(v)\,dv )g_k$, is a linear projection operator
that projects the standardized price-demand functions $X_t^\ast$ into
the subspace of $L^2[A,B]$ spanned by the orthonormal basis system
$B_K=\{g_1,\ldots,g_K\}$.

It is well known that a solution of the minimization problem (\ref
{opt3}) with respect to all $K$-dimensional orthonormal basis systems
$B_K$ can be determined by FPCA. This so-called ``best basis property''
of the empirical eigenfunctions $e_{T,1},\ldots,e_{T,K}$ is of
central importance for this paper; see Section~8.2.3 in \citet
{RamsayfdaBook2005}, among others. Note that the eigenvalues $\lambda
_{T,k}$ with $k\in\{1,\ldots,K\}$ may be of multiplicity $L>1$; in
this case $e_{T,k}\in E_k$ with $E_k=\myspan(e_{Tk,1},\ldots,e_{Tk,L})$.

A solution of (\ref{opt3}) is given by the set of eigenfunctions $\{
e_{T,1},\ldots,e_{T,K}\}$ that belong to the first $K$ greatest
eigenvalues $\lambda_{T,1}>\lambda_{T,2}>\cdots>\lambda_{T,K}>0$ of
the empirical covariance operator $\Gamma_T$ defined as
%
%
\begin{equation}
\label{Gamma1} (\Gamma_Tx) (u)=\int
_{A}^{B}\gamma_T(u,v)x(v)\,dv
\qquad\mbox{for all } x\in L^2[A,B],
\end{equation}
where the empirical covariance function $\gamma_T(u,v)$ is defined as
a local linear surface smoother in (\ref{covfun}). We use
this nonparametric version of $\gamma_T(u,v)$, since it can be applied
to the classical case of deterministic domains $\mathcal
{D}_t(X_t)=[A,B]$ as well as to the case of random domains $\mathcal
{D}_t(X_t)=[a_t,b_t]$ discussed in the following Section~\ref{rd}.
Contrary to this, the classical textbook definition of $\gamma_T(u,v)$
cannot be applied to the case of random domains.\footnote{The classical
definition is given by $\gamma_T(u,v)=T^{-1}\sum_{t=1}^TX_t(u)X_t(v)$.}
\subsection{Random domains $\mathcal{D}(X_t)=[a_t,b_t]$}\label{rd}

From a computational perspective, functional data observed on random
domains cause problems similar to sparsely observed functional data.
For the latter case there is already a broad stream of literature based
on the papers of \citet{Staniswal2010}, \citet{Yao2005}, \citet
{hall2006properties} and \citet{li2010uniform}.

We follow \citet{Yao2005} and compute the covariance function $\gamma
_T$ by local linear surface smoothing. Here, $\gamma_T(u,v)=\beta
_{T,0}$ and $\beta_{T,0}$ is determined by minimizing
%
%
\begin{eqnarray}
\label{covfun} &&\sum_{t=1}^T
\sum_{i,j=1}^{N_t}\kappa_2 \biggl(
\frac{(u_{ti}-u)}{b_\gamma},\frac{(u_{tj}-v)}{b_\gamma} \biggr
)\nonumber\\[-8pt]\\[-8pt]
&&\hspace*{13pt}\qquad{}\times\bigl\{X_t^\ast(u_{ti})X_t^\ast(u_{tj})-f
\bigl(\beta_T,(u,v),(u_{ti},u_{tj})\bigr)\bigr\}^2\nonumber
\end{eqnarray}
over $\beta_T=(\beta_{T,0},\beta_{T,11},\beta_{T,12})'\in\mathbb
{R}^3$, where $f(\beta_T,(u,v),(u_{ti},u_{tj}))=\beta_{T,0}+\break\beta
_{T,11}(u-u_{ti})+\beta_{T,12}(v-u_{tj})$, $u_{ti}$ are the observed
values of electricity demand, $b_\gamma$ is the smoothing parameter
that can be determined, for instance, by (generalized)
cross-validation, and $\kappa_2\dvtx \mathbb{R}^2\rightarrow\mathbb{R}$
is a two-dimensional kernel function such as the multiplicative kernel
$\kappa_2(x_1,x_2)=\kappa(x_1) \kappa(x_2)$ with $\kappa$ being a
standard univariate kernel such as the Epanechnikov kernel. See \citet
{Yao2005} and \citet{Fan1996} for further details.

\subsection{\texorpdfstring{The estimator $\{\hat{f}_1,\ldots,\hat{f}_K\}$}
{The estimator $\{f_1,\ldots,f_K\}$}}\label{defestim}

Given the analytic solution of the minimization problem (\ref{opt3}),
we can now define the estimator $\{\hat{f}_1,\ldots,\hat{f}_K\}$. The
only thing that we have to do is to replace the standardized
price-demand functions $X^\ast_t$ in (\ref{covfun}) by
their undersmoothed and standardized estimators $\tilde{X}^\ast_t$,
where $\tilde{X}^\ast_t=\tilde{X}_t/\|\hat{X}_t\|_2$.

Note that we scale the undersmoothed estimator $\tilde{X}_t$ with the
L2 norm of the estimator $\hat{X}_t$, which is optimally smoothed with
respect to the \emph{single} observation $X_t$. Undersmoothing of the
price-demand functions is always important if the target quantity, such
as the covariance function $\gamma_T(u,v)$, consists of an average
over all functions. The approximation of the norm $\|X_t\|$ does not
involve averages over all functions, such that we are better off to use
the norm of the classically smoothed curves $\|\hat{X}_t\|$.

Let us denote the estimator of the empirical covariance operator
$\Gamma_T$ by $\hat\Gamma_T$, defined as
%
%
\begin{equation}
\label{Gamma2} (\hat\Gamma_Tx) (u)=\int
_{A}^{B}\hat\gamma_T(u,v)x(v)\,dv
\qquad\mbox{for all } x\in L^2[A,B],
\end{equation}
where $\hat\gamma_T(u,v)$ is determined by minimizing equation (\ref
{covfun}) after replacing $X^\ast_t=X_t/\|X_t\|$ by $\tilde{X}^\ast
_t=\tilde{X}_t/\|\hat{X}_t\|_2$. Accordingly, we denote the first $K$
ordered eigenvalues and the corresponding eigenfunctions of $\hat
\Gamma_T$ by $\hat\lambda_{T,1}>\cdots>\hat\lambda_{T,K}$ and
$\hat{e}_{T,1},\ldots,\hat{e}_{T,K}$.

The estimator $\{\hat{f}_1,\ldots,\hat{f}_K\}$ is then defined as any
orthonormal rotation of the orthonormal basis system $\{\hat
{e}_{T,1},\ldots,\hat{e}_{T,K}\}$ determined by (\ref
{opt3}). The trivial case would be to use the empirical eigenfunctions
$\hat{e}_{T,1},\ldots,\hat{e}_{T,K}$ directly as basis functions such
that $\hat{f}_k=\hat{e}_{T,k}$ for all $k\in\{1,\ldots,K\}$. It is
generally left to the statistician to choose an appropriate orthonormal
rotation scheme, which facilitates the interpretation. In our
application we use the well-known VARIMAX-rotation.

Following our assumptions on the data generating process in
(\ref{fm}), we use the basis system $\{\hat{f}_1,\ldots,\hat
{f}_K\}$ in order to re-estimate the functions $X_1,\ldots,X_T$ by
%
%
\begin{equation}
\label{approx} \hat{X}_t^f=\sum
_{k=1}^K\hat\beta_{tk}
\hat{f}_k,
\end{equation}
where the parameters $\hat\beta_{tk}$ are defined according to
(\ref{OLS}). This is a crucial step of the estimation
procedure. Given that our model assumption in (\ref{fm}) is
true, the original single cubic smoothing splines estimates $\hat
{X}_t$ will be much less efficient estimators of the price-demand
functions $X_t$ than the estimators $\hat{X}^f_t$ since the latter use
the information of the whole data set.
\subsection{A note on convergence}\label{NoteonConvergence}
Assume that we are able to observe the (unobservable) set of functions
$\{X_1,\ldots,X_T\}$ as defined in (\ref{fm}) but with
deterministic domains $\mathcal{D}(X_t)=[A,B]$. In this case the $K$
empirical eigenfunctions $e_{T,1},\ldots,e_{T,K}$ can be determined
from the empirical covariance operator $\Gamma_T$ as defined in
(\ref{Gamma1}) based on the classical definition of the
empirical covariance function $\gamma_T(u,v)=K^{-1}\sum
_{k=1}^KX_{t_k}(u)X_{t_k}(v)$. Actually, only a subset of at least $K$
linear independent functions, say, $X_{t_1},\ldots,X_{t_K}$, would
suffice to determine the $K$ empirical eigenfunctions $e_{T,1},\ldots,e_{T,K}$.

In this case, the determination of the basis system
$\{e_{T,1},\ldots,e_{T,K}\}$ is a mere algebraic problem. Furthermore,
the space $\mathcal{H}_K$ spanned by the basis system
$\{e_{T,1},\ldots,e_{T,K}\} $ does not depend on the data. By the
definition in (\ref {fm}), two sets of functions $\{X_1,\ldots,X_T\}$
and $\{X_1,\ldots,X_{T'}\}$ span the same space $\mathcal{H}_K$ for all
$T'\geq T\geq K$, such that also the corresponding basis systems
$\{e_{T,1},\ldots,e_{T,K}\}$ and $\{e_{T',1},\ldots,e_{T',K}\}$ span the
same space $\mathcal{H}_K$.

Note that we do not observe the functions $X_t$ but the noisy
discretization points $y_{th}=X_t(u_{th})+\varepsilon_{th}$. Starting
with the first scenario of deterministic domains $\mathcal
{D}(X_t)=[A,B]$, the determination of the estimated basis system $\{
\hat{e}_{T,1},\ldots,\hat{e}_{T,K}\}$ can be done from the estimated
empirical covariance operator $\hat\Gamma_T$ defined in
(\ref{Gamma2}). Again, this on its own is a mere algebraic
problem but it yields to our consistency argument.

The estimated eigenfunction $\hat{e}_{Tk}$ can be written as a
continuous function of $\hat{X}_{1},\ldots,\hat{X}_{T}$, say, $\hat
{e}_{Tk}=g_k(\hat{X}_{1},\ldots,\hat{X}_{T})$. By the
continuous\vspace*{1pt}
mapping theorem, $\hat{e}_{Tk}=g_k(\hat{X}_{1},\ldots,\hat{X}_{T})$
converges to $e_{Tk}=g_k(X_{1},\ldots,X_{T})$ as $\hat
{X}_{t}(u)\rightarrow X_{t}(u)$, for example, in probability as
$N_t\rightarrow\infty$ for all $u\in[A,B]$ and $t\in\{1,\ldots,T\}$
with $T\geq K$. We additionally have to assume that all involved
smoothing parameters go against zero appropriately fast such that
$N\underbar{b}_K\rightarrow\infty$, $Nb^{\mathrm{opt}}\rightarrow
\infty$, and $NTb_\gamma\rightarrow\infty$ [\citet
{benedetti1977nonparametric}]. Eigenfunctions $e_k$ are determined up
to sign changes and it is assumed that the correct signs are chosen.

In this sense we can state that $\hat{\mathcal{H}}_K=\myspan(\hat
{e}_{T,1},\ldots,\hat{e}_{T,K})$ converges to $\mathcal{H}_K=\myspan
(e_{T,1},\ldots,e_{T,K})$, which is all we can achieve for (functional)
factor models, since the single factors $f_k$ remain unidentifiable.

Finally, it only remains to consider the scenario of random
domains\break
$\mathcal{D}(X_t)=[a_t,b_t]$. Also, in this case any two points
$(u,v)\in[A,B]$ have to be covered by at least $K$ price-demand
functions, which are fulfilled asymptotically. By our assumptions the
time series $(a_t)$ and $(b_t)$ are processes with $A\leq a_t<b_t\leq
B$ and the marginal p.d.f.s of $a_t$ and $b_t$ are given by $f_a(z_a)>0$
and $f_b(z_b)>0$ for all $z_a,z_b\in[A,B]$ and $t\in\{1,\ldots,T\}$.
This yields that
\[
\operatorname{Pr} \bigl(a_t\in[A,A+\varepsilon] \bigr)>0
\quad\mbox{and}\quad
\operatorname{Pr} \bigl(b_t\in[B-\varepsilon,B] \bigr)>0 \qquad\mbox{for any }
\varepsilon>0,
\]
such that for $T\rightarrow\infty$ with probability one, there are
sub-series $(a_s)$ and $(b_s)$ of $(a_t)$ and $(b_t)$ for which the
boundary points $A$ and $B$ are accumulation points. From this it
follows that as $T\rightarrow\infty$ we can find always more than $K$
functions $X_t$ that cover the points $u,v\in[A,B]$.

To conclude, consistency of our estimation procedure relies on our
model assumptions in (\ref{fm0}) and (\ref{fm}) and is
driven by the consistency of the first step estimators of the
price-demand functions $X_t(u)$.

\section{Application}\label{a}
In this section we apply our estimation procedure of the FFM described
in Section~\ref{e} to the data set described in Section~\ref{d}. In
Section~\ref{ai} we show how to interpret the factors and demonstrate
an exemplary analysis of the scores and in Section~\ref{v} we validate
the crucial model assumptions.

A drawback of the cross-validation criterion in (\ref{CVEq})
is that it depends on the unknown dimension $K$. Therefore, first, we
determine optimal undersmoothing parameters $\underbar{b}_K$ for
several values of $K$ and, second, choose the dimension $K$, which
minimizes the AIC of \citet{Yao2005}.

The AIC type criterion indicates an optimal dimension of $K=2$ (AIC
values in Table~\ref{CVKTable} are shown as differences from the
lowest AIC value). These first two basis functions are able to explain
$99.95\%$ of the variance. The minimization of the cross-validation
criterion (\ref{CVEq}) for $K=2$ yields an undersmoothing parameter
of $\underbar{b}_{ 2}$ that is only two-tenths of the usual
cross-validation smoothing parameter $b^{\mathrm{opt}}$; see Table~\ref{CVKTable}.

%
\begin{table}[b]
\tablewidth=240pt
\caption{Undersmoothing parameters $\protect\underbar{b}_K$ (shown as
fractions of the usual cross-validation smoothing parameter $b^{\mathrm{opt}}$),
AIC values (shown as differences from the lowest AIC value) and
cumulative variances for the dimensions $K\in\{1,2,3\}$}
\label{CVKTable}
\begin{tabular*}{\tablewidth}{@{\extracolsep{\fill}}lcd{3.1}c@{}}
\hline
$\bolds{K}$ & $\bolds{\underbar{b}_K/b^{\mathrm{opt}}}$
& \multicolumn{1}{c}{\textbf{AIC}} & \textbf{Cum. var.}\\
\hline
$1$&$0.1$&596.4&$92.62\%$\\
$2$&$0.2$&0&\emph{$99.95\%$}\\
$3$&$0.3$&52.9& $99.97\%$\\
\hline
\end{tabular*}
\end{table}
%

\begin{figure}

\includegraphics{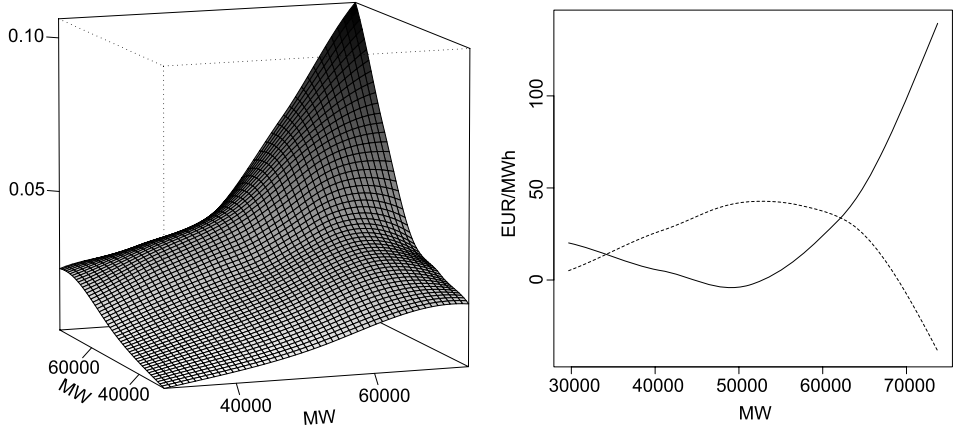}

\caption{\textsc{Left panel}: Empirical covariance function $\hat\gamma
_T$. \textsc{Right panel}: VARIMAX rotated basis functions $\hat{f}_{1}$
(solid line) and $\hat{f}_{2}$ (dashed line), scaled by the average
scores $\bar{\hat\beta}_{\cdot1}$ and $\bar{\hat\beta}_{\cdot2}$, respectively.}
\label{Fig4}
\end{figure}

Based on the undersmoothed and scaled estimators $\tilde{X}^\ast
_t=\tilde{X}_t/\|\hat{X}_t\|$, we compute the estimator $\hat{\gamma
}_T$ of the empirical covariance function $\gamma_T$ by local linear
surface smoothing, as explained in Section~\ref{defestim}. The result
is shown in the left panel of Figure~\ref{Fig4}. The plot of the
estimator $\hat{\gamma}_T$ shows clearly that the sample variance of
the standardized price-demand functions $\tilde{X}^\ast_t$ increases
monotonically with electricity demand.

%
%

The estimators of the first two empirical eigenfunctions $\hat
{e}_{T,1}$ and $\hat{e}_{T,2}$ are determined from the
eigendecomposition of a discretized version of the estimated empirical
covariance function $\hat{\gamma}_T$ using an equidistant grid of
$n\times n$ discretization points of the plane $[A,B]^2$. The
estimation of the smooth eigenfunctions by discretizing the smooth
covariance function is common in the FDA literature; see, for example,
\citet{Rice1991}.

In order to find an appropriate number $n$ of discretization points,
there is the following trade-off, which has to be considered: on the
one hand, $n$ must be small enough that the algorithm to compute the
eigendecomposition runs stable. On the other hand, $n$ must be great
enough that the $n\times n$-matrix of discretization points forms a
good approximation to the covariance function. The choice of $n=50$
appears to be appropriate for our application. As a robustness check we
also tried values of $n$ ranging from $20$ to $70$, which yield nearly
identical results.

We rotate the basis system of the estimated eigenfunctions $\{\hat
{e}_{T,1},\hat{e}_{T,2}\}$ by the VARIMAX-criterion in order to get
interpretable basis functions $\hat{f}_1$ and $\hat{f}_2$. The two
rotated basis functions $\hat{f}_1$ and $\hat{f}_2$ explain $58.63\%$
and $41.32\%$ of the total sample variance of the price-demand
functions $\hat{X}_t$.

It is convenient to choose an appropriate scaling of the graphs of the
basis functions $\hat{f}_1$ and $\hat{f}_2$ in order to plot them
with a reasonable order of magnitude. We scale the graphs by their
corresponding average scores $\bar{\hat\beta}_{\cdot i}=T^{-1}\sum
_{t=1}^T\hat\beta_{ti}$ for $i\in\{1,2\}$. In the right panel of
Figure~\ref{Fig4} the graph of $\hat{f}_1 \bar{\hat\beta}_{\cdot 1}$
is plotted as a solid line, whereas the graph of $\hat{f}_2 \bar{\hat
\beta}_{\cdot 2}$ is plotted as a dashed line.

Given the basis system $\{\hat{f}_1,\hat{f}_2\}$, we re-estimate the
functions $X_1,\ldots,X_T$ by (\ref{approx}) such that
\[
\hat{X}_t^f=\hat\beta_{t1}
\hat{f}_1+\hat\beta_{t2}\hat{f}_2.
\]
To simplify the notation, we write $\hat{X}_t=\hat{X}^f_t$ from now
on. The coefficients $\hat\beta_{t1}$ and $\hat\beta_{t2}$ are
determined by OLS regressions of $\hat{X}_t$ simultaneously on
%
%
\begin{figure}

\includegraphics{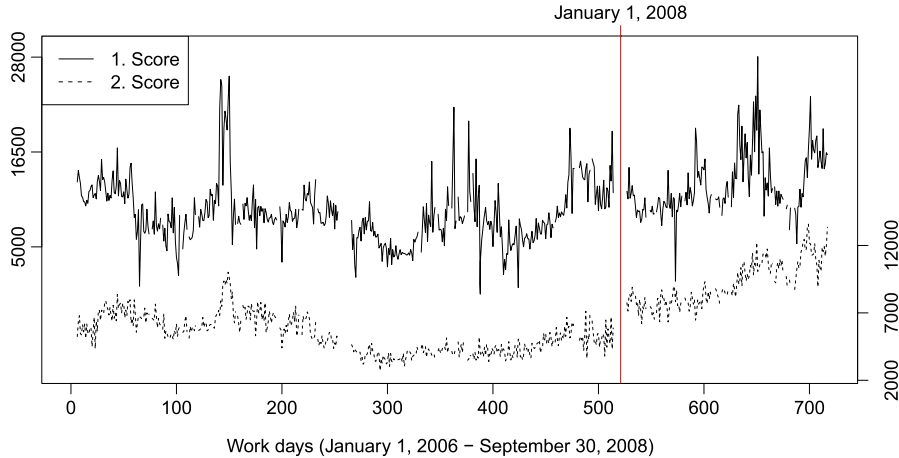}

\caption{Time series of the first scores $(\hat{\beta}_{t1})$ (solid
line) and second scores $(\hat{\beta}_{t2})$ (dashed line). The
vertical red line separates the initial learning sample from the
initial forecasting sample. Gaps in the time series correspond to holidays.}
\label{FigScores}
\end{figure}
%
%
$\hat{f}_1$ and $\hat{f}_2$ after discretizing the functions at the $N_t$
observed values of electricity demand $u_{t1},\ldots,u_{tN_t}$. The
time series of the scores are shown in Figure~\ref{FigScores}.

\subsection{Interpretation of the factors and exemplary analysis of
the scores}\label{ai}
Remember that we do not use a mean function in our FFM. Consequently,
the classical interpretation of the factors $\hat{f}_k$ as
perturbations of the mean does not apply. A reasonable interpretation
of the estimated factors $\hat{f}_{1}$ and $\hat{f}_{2}$ can be
derived from the classical micro-economic point of view on electricity
spot prices; see also the discussion in Section~\ref{d}.

This point of view allows us to interpret the price-demand functions
$\hat{X}_t(u)=\hat\beta_{t1}\hat{f}_1(u)+\hat\beta_{t2}\hat
{f}_2(u)$ as daily \emph{empirical merit order curves} or \emph
{empirical supply functions}, where the shape of the curves $\hat
{X}_t$ is determined by the factors $\hat{f}_1$ and $\hat{f}_2$ and
the scores $\hat\beta_{t1}$ and $\hat\beta_{t2}$. For example,
steep empirical supply functions have high score ratios $\hat\beta
_{t1}/\hat\beta_{t2}$ and vice versa. Since steep supply functions
are associated with high prices, we could interpret the first factor
$\hat{f}_1$ as the high-price component and the second factor $\hat
{f}_2$ as the moderate-price component. In general, any interpretation
of the factors has to be done with caution since they are only
identified up to orthonormal rotations.

Particularly, the scores $\hat\beta_{t1}$ and $\hat\beta_{t2}$ are
useful for a further analysis of the dynamics of the empirical supply
functions. For example, researchers or risk analysts, who wish to
predict days with high electricity prices, could try to predict days
with steep empirical supply functions $\hat{X}_t$.

Days with steep supply functions represent market situations with
capacity constraints, that is, situations in which power plants with
high generation costs are needed to supply the demanded amount of
electricity. There are several causes for capacity constraints, such as
extreme temperatures or power plant outages.

%
\begin{figure}

\includegraphics{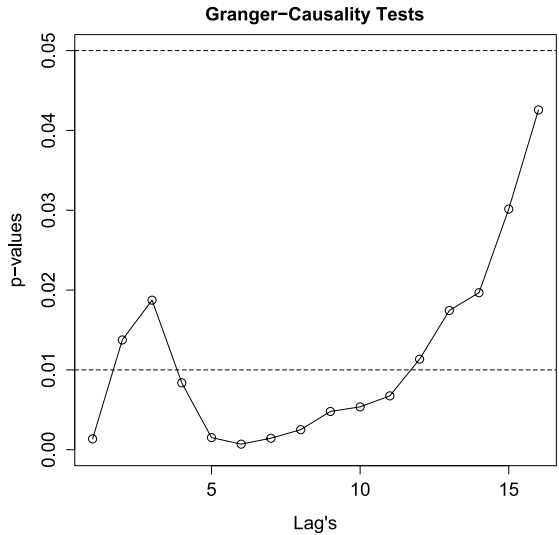}

\caption{$p$-values of Granger-causality tests of whether the
time-varying steepness of the price-demand functions [quantified as
time series of score ratios ($\hat\beta_{t1}/\hat\beta_{t2}$)] is
Granger-caused by past values of the time series of extreme temperatures.}
\label{Granger}
\end{figure}
%

In fact, the time-varying steepness of the empirical supply functions
[quantified as time series of score ratios $(\hat\beta_{t1}/\hat
\beta_{t2})$] is Granger-caused by the time series of extreme
temperatures (defined as absolute temperature deviations from the mean
temperature), where the temperature data is available from the German
Weather Service (\href{http://www.dwd.de}{www.dwd.de}). Figure~\ref{Granger} shows the
$p$-values of the corresponding Granger-causality tests [\citet
{granger1969investigating}].

\subsection{Validation of the model assumptions}\label{v}
The overall in-sample data fit of the estimated spot prices $\hat
{y}_{th}=\hat{X}_t(u_{th})$, measured by the $R^2$-parameter, is given
by $R^2=0.92$ and indicates a good model fit. Nevertheless, our
implicit stability assumption in (\ref{fm}) that $X_t\in\mathcal
{H}_K$ for all $t\in\{1,\ldots,T\}$, that is, that all functions $X_t$
are elements of the same space $\mathcal{H}_K$, may be seen as critical.

In our context it is impossible to validate the stability assumption by
statistical tests such as in \citet{Benko2009a} since we do not assume
that the time series of the scores $(\beta_{t1})$ and $(\beta_{t2})$
are stationary. However, we can compare different basis systems
estimated from subsets of the data with each other. The stability
assumption can be seen as supported if all of these subset-basis
functions span the same space $\hat{\mathcal{H}}_K$.

We define half-yearly data-subsets 6-1, 6-2, 7-1, 7-2 and one
nine-month data-subset 8-1 by choosing according to subsets of the
index set $\{1,\ldots,T\}$ and investigate the $R^2$-parameters from
subset regressions---such as, for example,
$\hat{e}^{(6\mbox{-}1)}_{1}(u)$ simultaneously on
$\hat{e}^{(6\mbox{-}2)}_{1}(u)$ and $\hat{e}^{(6\mbox{-}2)}_{2}(u)$.
This assesses whether the eigenfunction $\hat{e}^{(6\mbox{-}1)}_{1}$
can be seen as an element of the space spanned by the basis system
$\{\hat{e}^{(6\mbox{-}2)}_{1},\hat{e}^{(6\mbox{-}2)}_{2}\}$.

The results are given in Table~\ref{tab1} and clearly support our
assumption that $X_t\in\mathcal{H}_K$ for all $t\in\{1,\ldots,T\}$.
The $R^2$-values with respect to the first eigenfunctions $\hat
{e}^{(6\mbox{-}1)}_1,\hat{e}^{(6\mbox{-}2)}_1,\ldots,\hat
{e}^{(8\mbox{-}1)}_1$ and $\hat{e}_{T,1}$ are all greater than or
equal to $0.99$. Also, the $R^2$-values with respect to the second
eigenfunctions $\hat{e}^{(6\mbox{-}1)}_2,\hat{e}^{(
6\mbox{-}2)}_2,\ldots,\hat{e}^{(8\mbox{-}1)}_2$ and $\hat{e}_{T,2}$
indicate no clear violation of our model assumption.

%
\begin{table}
\tabcolsep=4pt
\caption{Descriptive validation of the assumption that $X_t\in
\mathcal{H}_K$ for all $t\in\{1,\ldots,T\}$. The list elements are
$R^2$-values, which stem from subset regressions of, for example, the
eigenfunction $\hat{e}^{(6\mbox{-}1)}_{1}$ on the eigenfunctions $\{
\hat{e}^{(6\mbox{-}2)}_{1},\hat{e}^{(6\mbox{-}2)}_{2}\}$ in the
upper left case with $R^2=0.99$}
\label{tab1}
\begin{tabular*}{\tablewidth}{@{\extracolsep{\fill}}lccccc@{}}
\hline
&$\bolds{\hat{e}^{(6\mbox{-}1)}_{1}}$ $\bolds{\hat{e}^{(
6\mbox{-}1)}_{2}}$&$\bolds{\hat{e}^{(6\mbox{-}2)}_{1}}$ $\bolds{\hat
{e}^{(6\mbox{-}2)}_{2}}$&$\bolds{\hat{e}^{(7\mbox{-}1)}_{1}}$
$\bolds{\hat{e}^{(7\mbox{-}1)}_{2}}$&
$\bolds{\hat{e}^{(7\mbox{-}2)}_{1}}$ $\bolds{\hat{e}^{(7\mbox{-}2)}_{2}}$&$\bolds{\hat
{e}_{1}^{(8\mbox{-}1)}}$ $\bolds{\hat{e}^{(8\mbox{-}1)}_{2}}$\\
\hline
$\{\hat{e}^{(6\mbox{-}1)}_{1},\hat{e}^{(6\mbox{-}1)}_{2}\}$&---
--- &0.99 0.95&1.00 0.96&1.00 0.89&1.00 0.99\\[2pt]
$\{\hat{e}^{(6\mbox{-}2)}_{1},\hat{e}^{(6\mbox{-}2)}_{2}\}
$&0.99 0.95&--- --- &0.99 0.83&1.00 0.98&1.00 0.95\\[2pt]
$\{\hat{e}^{(7\mbox{-}1)}_{1},\hat{e}^{(7\mbox{-}1)}_{2}\}
$&1.00 0.96&0.99 0.83&--- --- &1.00 0.78&1.00 0.95\\[2pt]
$\{\hat{e}^{(7\mbox{-}2)}_{1},\hat{e}^{(7\mbox{-}2)}_{2}\}
$&1.00 0.89&1.00 0.98&1.00 0.78&--- --- &1.00 0.89\\[2pt]
$\{\hat{e}^{(8\mbox{-}1)}_{1},\hat{e}^{(8\mbox{-}1)}_{2}\}
$&1.00 0.99&1.00 0.95&1.00 0.95&1.00 0.89&--- --- \\
[6pt]
$\{\hat{e}_{T1},\hat{e}_{T2}\}$ &1.00 0.99&1.00 0.96&1.00 0.95&1.00
0.90&1.00 0.99\\
\hline
\end{tabular*}
\end{table}

The $R^2$-values with respect to the second eigenfunctions are
systematically smaller than those with respect to the first
eigenfunctions, since the first order bias term of an estimated
eigenfunction is inversely related to the pairwise distances of its
eigenvalue to all other eigenvalues; see \citet{Benko2009a}, Theorem
2(iii). By construction, these distances are greatest for the first
eigenvalue.

Finally, we test for (non-)stationarity of the time series of the
scores $(\hat\beta_{t1})$ and $(\hat\beta_{t2})$ using the usual
testing procedures such as the KPSS-tests for stationarity and
ADF-tests for nonstationarity (with a 5\%-significance level for all
tests). The results allow us to assume that the time series of the
scores $(\hat\beta_{t1})$ and $(\hat\beta_{t2})$ are nonstationary.
Detailed reports are not shown for reasons of space but can be
reproduced using the R-Code provided as part of the supplementary
material; see \citet{liebl2013}.

This section demonstrates a very good and stable in-sample fit of our
FFM. Of course, this cannot guarantee a good out-of-sample performance.

\section{Forecasting}\label{f}
For our forecasting study we divide the data set into a learning sample
of days $t\in\{1,\ldots,T_L\}$ and a forecasting sample of days $t\in
\{T_L+1,\ldots,T\}$, where the initial $T_L+1$ corresponds to January
1, 2008 and $T$ to September 30, 2008. The learning sample is used to
estimate the parameters and the forecasting sample is used to assess
the forecast performance. We enlarge the learning sample after each
$\ell$ days ahead forecast by one day. For $\ell\in\{1,\ldots,20\}$
days ahead forecasts this leads to $T-T_L-(\ell-1)=717-521-(\ell
-1)=197-\ell$ work days that can be used to assess the forecast
performance of our model.

\begin{figure}

\includegraphics{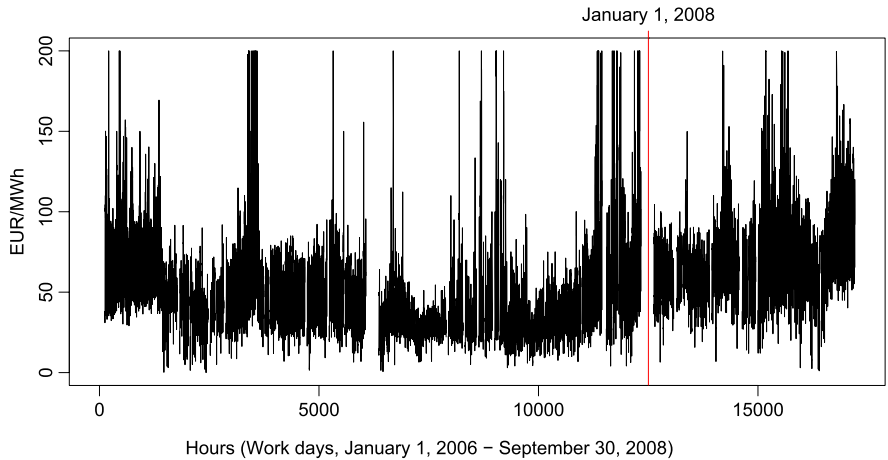}

\caption{The whole data set of $17\mbox{,}208$ hourly electricity spot
prices. The vertical red line separates the initial learning sample
from the initial forecasting sample. Gaps in the time series correspond
to holidays.}
\label{FigSpotPrices}
\end{figure}

Figure~\ref{FigSpotPrices} shows the whole data set of $717\cdot
24=17\mbox{,}208$ hourly electricity spot prices along with the indicated time
spans of the initial learning and forecasting sample. Gaps in the time
series correspond to holidays. At least from a visual perspective, the
learning sample and the forecasting sample are of comparable complexity.

%
%

In the following Section~\ref{fFFM} we discuss forecasting of
electricity spot prices using the FFM. In Section~\ref{fc} we formally
introduce four competing forecasting models (two classical and two FDA
models), and in Section~\ref{fp} we compare their predictive performance.

As noted above for the FFM, the two competing FDA models also use
learning data with electricity spot prices below 200 EUR/MWh only. In
contrast, the forecasting sample retains all data, including those with
spot prices above 200 EUR/MWh. Advanced outlier forecast procedures,
which might yield better predictive performances, are beyond the scope
of this paper.

\subsection{Forecasting with the FFM}\label{fFFM}
The computation of the $\ell$ days ahead forecast $\hat
{y}_{T_L,h}(\ell)\in\mathbb{R}$ of the electricity spot price
$y_{T_L+\ell,h}$ given the information set of the learning sample,
say, $\mathcal{I}_{T_L}$, involves the computation of the conditional
expectation of a nonlinearly transformed random variable, namely,
$E[f_k(u_{T_L+\ell,h})|\mathcal{I}_{T_L}]$. We approximate the latter
using the naive plug-in predictor $\hat{f}_k(\hat{u}_{T_L,h}(\ell
))$, where $\hat{u}_{T_L,h}(\ell)=E[u_{T_L+\ell,h}|\mathcal
{I}_{T_L}]$. This yields to
%
%
\begin{equation}
\label{fcst}
\hat{y}_{T_L,h}(\ell)=\hat\beta_{T_L,1}(\ell)
\hat{f}_1 \bigl(\hat{u}_{T_L,h}(\ell) \bigr)+\hat
\beta_{T_L,2}(\ell)\hat{f}_2 \bigl(\hat{u}_{T_L,h}(
\ell) \bigr),
\end{equation}
where $\hat\beta_{T_L,1}(\ell)$, $\hat\beta_{T_L,2}(\ell)$ and
$\hat{u}_{T_L,h}(\ell)$ are the $\ell$ days ahead forecasts of the
scores $\hat\beta_{T_L+\ell,1}$, $\hat\beta_{T_L+\ell,2}$ and of
the electricity demand value $u_{T_L+\ell,h}$.

The naive plug-in predictor $\hat{f}_k(\hat{u}_{T_L,h}(\ell))$ is a
rather simple approximation of the conditional expectation
$E[f_k(u_{T_L+\ell,h})|\mathcal{I}_{T_L}]$. Here, it performs very
well because the basis functions are relatively smooth. In the case of
more complex basis functions, it might be necessary to improve the
approximation using higher order Taylor expansions of $\hat{f}_k$
around $\hat{u}_{T_L,h}(\ell)$.

We use the following univariate $\operatorname{SARIMA}(0,1,6)\times
(0,1,1)_5$-models to forecast the time series of the scores $(\hat
\beta_{ti})$ with $i\in\{1,2\}$:
%
%
\begin{equation}
\label{SARIMA12} (1-B ) \bigl(1-B^5 \bigr)\hat
\beta_{ti} = \Biggl(1+\sum_{l=1}^6
\delta_{il} B^l \Biggr) \bigl(1+\delta^S_{i}
B^{5} \bigr) \omega_{ti},
\end{equation}
where $B$ is the back shift operator. In order to ensure that the
SARIMA models (\ref{SARIMA12}) are not sample dependent, we select
them from a set of reasonable alternative SARIMA models, where all of
them are confirmed by the usual diagnostics on the residuals. Each of
the confirmed models is applied to different subsets of the learning
sample and the final model selection is done by the AIC.\footnote{The
interested reader is referred to the R-Code provided as part of the
supplementary material; see \citet{liebl2013}.} As usual, the $\ell$
days ahead forecasts $\hat\beta_{T_L,1}(\ell)$ and $\hat\beta
_{T_L,2}(\ell)$ are given by the conditional expectations of $\hat
\beta_{T_L+\ell,1}$ and $\hat\beta_{T_L+\ell,2}$ given the data
from the learning sample; see, for example, \citet{Brockwell1991}.

%
\begin{figure}

\includegraphics{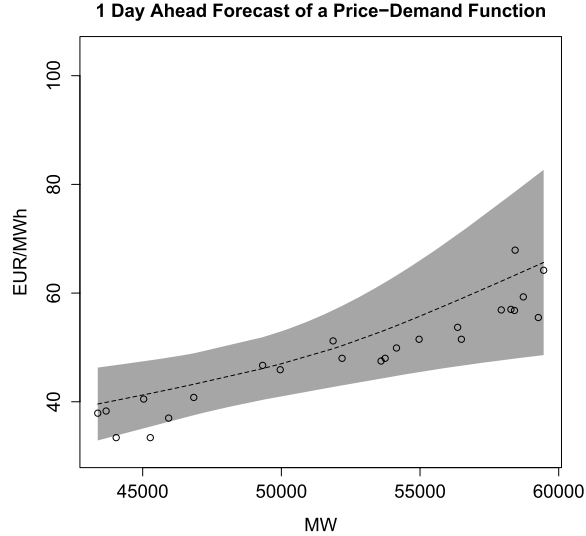}

\caption{A comparison of the 24 hourly electricity spot prices
$y_{th}$ (circles) and the $1$ day ahead forecast of the price-demand
function $X_t$ (dashed line) of January 25, 2008. The 95\% forecast
interval is plotted as a gray shaded band.}
\label{FigForecastPriceDemandFCTs}
\end{figure}
%

A first visual impression of the forecast performance is given in
Figure~\ref{FigForecastPriceDemandFCTs}, which compares the $24$
hourly spot prices $y_{th}$ with the $1$ day ahead forecast of the
price-demand function $X_t$. The $1$ day ahead forecast of the
price-demand function $X_t$ is defined as
%
%
\begin{equation}
\label{FcstPDFct} \hat{X}_{T_L}(\ell)=\hat
\beta_{T_L,1}(\ell)\hat{f}_1+\hat\beta_{T_L,2}(\ell)
\hat{f}_2 \in L^2[A,B].
\end{equation}
Additionally, a 95\% forecast interval is plotted as a gray shaded
band. The forecast interval is computed on the basis of the 95\%
forecast intervals of the SARIMA forecasts $\hat\beta_{T_L,1}(\ell)$
and $\hat\beta_{T_L,2}(\ell)$ and has to be interpreted as a
conditional forecast interval given the realizations $\hat{f}_1$ and
$\hat{f}_2$.

In order to be able to forecast the hourly electricity spot prices
$y_{th}$, we also have to forecast the hourly values of electricity
demand $u_{th}$; see equation (\ref{fcst}). Given our definition of
electricity demand in Section~\ref{d}, a $\ell$ days ahead forecast
of electricity demand $u_{T_L,h}(\ell)$ involves forecasting gross
demand for electricity as well as wind power infeed data. The
statistician has to choose appropriate models---one for gross
electricity demand such as that proposed in \citet{Antoch} and another
for wind power such as that proposed in \citet{Lau2010}. For the sake
of simplicity, we use the two reference cases of a ``persistence'' and
an ``ideal'' forecast of electricity demand:
\begin{longlist}
\item[\textit{persistence}] The persistence (or ``no-change'') forecast $\hat
{u}^{\mathrm{persi}}_{T_L,h}(\ell)$ is given by the last value of electricity
demand that is still within the learning sample, that is, $\hat
{u}^{\mathrm{persi}}_{T_L,h}(\ell)=u_{T_L,h}$.
\item[\textit{ideal}] The ideal forecast is given by $u_{T_L+\ell,h}$ itself,
that is, $\hat{u}^{\mathrm{ideal}}_{T_L,h}(\ell)=u_{T_L+\ell,h}$.
\end{longlist}
This yields a range for possible electricity demand forecasts with
bounds that can be easily interpreted.

%
\begin{figure}

\includegraphics{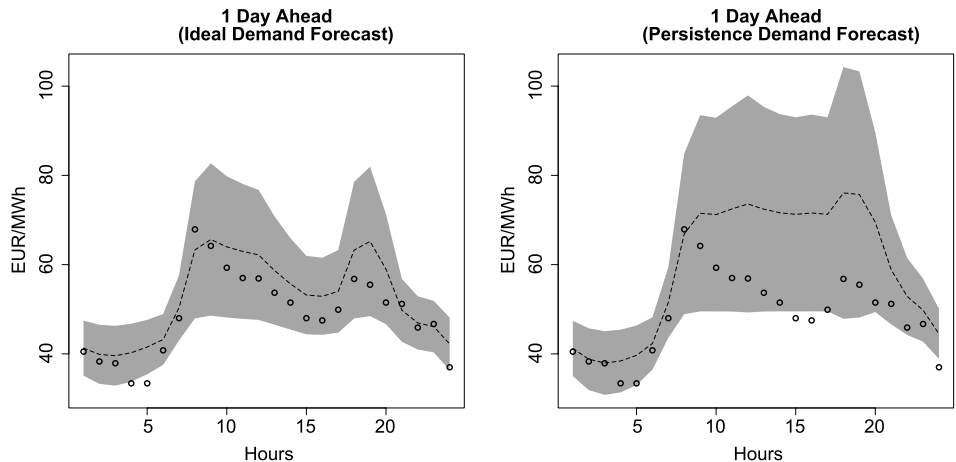}

\caption{\textsc{Left panel}: comparison of the spot prices $y_{th}$
(circles) and the $1$ day ahead forecasts $\hat{y}_{T_L,h}(\ell)$
(dashed line) of January 25, 2008 based on ideal demand forecasts.
\textsc{Right panel}: comparison of the spot prices $y_{th}$ (circles) and the
$1$ day ahead forecasts $\hat{y}_{T_L,h}(\ell)$ (dashed line) of
January 25, 2008 based on persistence demand forecasts. \textsc{Both
panels}: the 95\% forecast intervals are plotted as gray shaded bands.}
\label{Forecast1aheadhourlyorder}
\end{figure}
%

A first visual comparison of the observed hourly electricity spot
prices $y_{T_L+1,h}$ with their $1$ day ahead forecasts $\hat
{y}_{T_L,h}(1)$ is given in Figure~\ref{Forecast1aheadhourlyorder}.
The left panel demonstrates the ideal forecast case and shows the spot
prices $y_{th}$ (circles) and their $1$ day ahead forecasts $\hat
{y}_{T_L,h}(1)$ (dotted line) based on the electricity demand forecasts
$\hat{u}^{\mathrm{ideal}}_{T_L,h}(1)$. The right panel demonstrates the
persistence case and shows the spot prices $y_{th}$ (circles) and their
$1$ day ahead forecasts $\hat{y}_{T_L,h}(1)$ (dotted line) based on
the electricity demand forecasts $\hat{u}^{\mathrm{persi}}_{T_L,h}(1)$. The
95\% forecast intervals are plotted as gray shaded bands. The forecast
interval shown in the right panel is much broader than that shown in
the left panel. This is because the forecasted electricity spot prices
based on the persistence electricity demand forecasts are too high, and
higher electricity spot prices have broader 95\% forecast intervals;
see Figure~\ref{FigForecastPriceDemandFCTs}.\looseness=1

\subsection{Competing forecast models}\label{fc}
In this section we introduce the four competing forecast models (two
classical and two FDA models). The two classical models, referred to as
AR and MR models, are archetypal representatives of the classical
approaches in the literature on forecasting electricity spot prices;
see, for example, \citet{Kosater2006}. The AR model is an
autoregressive model and the MR model is the Markov regime switch model
for electricity spot prices proposed by \citet{HuismanRAndDeJong2003}.

The two FDA models are the above-discussed DSFM model of \citet
{Park2009} and the semi-functional partial linear (SFPL) model of \citet
{vilar2012forecasting}. Both of these FDA models have been successfully
applied to forecast electricity spot prices [\citet{Haerdle2010} and
\citet{vilar2012forecasting}] and are expected to be more challenging
competitors for our FFM than the two classical models.

Before the formal introduction of the four alternative forecast models,
we need some unifying notation. The problem is that the two classical
models, AR and MR, are designed to forecast only daily \emph
{aggregated} peakload and baseload spot prices defined as
\[
y_t^P=\log\Biggl(\frac{1}{12}\sum
_{h=9}^{20}y_{th} \Biggr) \quad\mbox{and}\quad
y_t^B = \log\Biggl(\frac{1}{24}\sum
_{h=1}^{24}y_{th} \Biggr).
\]

In contrast to this, the three FDA models are designed to forecast the
\emph{hourly} electricity spot prices $y_{th}$. Therefore, we define
the forecasts of the peakload-aggregates $\hat{y}_{T_L}^P(\ell
|\mathtt{Model})$ and baseload-aggregates $y_{T_L}^B(\ell|\mathtt
{Model})$ of the FDA models as
\[
\hat{y}_{T_L}^P(\ell|\mathtt{Model})=\log\Biggl(
\frac{1}{12}\sum_{h=9}^{20}
\hat{y}_{T_L,h}(\ell|\mathtt{Model}) \Biggr)
\]
and
\[
\hat{y}_{T_L}^B(\ell|\mathtt{Model})=\log\Biggl(
\frac{1}{24}\sum_{h=1}^{24}
\hat{y}_{T_L,h}(\ell|\mathtt{Model}) \Biggr),
\]
where $\hat{y}_{T_L,h}(\ell|\mathtt{Model})$ is the $\ell$ days
ahead hourly electricity spot price forecast of the $\mathtt{Model}\in
\{\mathrm{FFM}, \mathrm{DSFM}, \mathrm{SFPL}\}$. By Jensen's
inequality, these definitions yield aggregated forecasts of the FDA
models, which tend to be too high, that is, $E[y_{T_L+\ell}^A|\mathcal
{I}_{T_L}]\leq\hat{y}_{T_L}^A(\ell|\mathtt{Model})$ with $A\in\{
P,B\}$. Therefore, the RMSEs of the FDA models shown in Figure~\ref{RMSE} tend to be inflated and can be interpreted as being conservative.

In the following we formally introduce the four competing forecast
models. Further details can be found in \citet{Kosater2006}, \citet
{Park2009} and \citet{vilar2012forecasting}.

\subsubsection*{AR} The first benchmark model is the classical
AR(1) model with an additive constant drift component and a
time-varying deterministic component. The AR model can be defined as
%
%
\begin{equation}
\label{m1} y_t^A=d^A+g^A_t+
\alpha y^A_{t-1}+\omega^A_t,\qquad
\omega_t^A\sim\mathcal{N}\bigl(0,\sigma^2_{\omega^A}
\bigr),
\end{equation}
where $A\in\{P,B\}$ refers to the type of aggregation (peakload or
baseload), $d^A$ is the constant drift parameter, and $g^A_t$ captures
daily, weekly and yearly deterministic effects of the peakload and
baseload prices, respectively.

\subsubsection*{MR} The second benchmark model is the Markov
regime switch model proposed by \citet{HuismanRAndDeJong2003}. The MR
model extends the AR model (\ref{m1}) and distinguishes between two
different regimes $R_t^A\in\{M,S\}$, where $M$ denotes the regime of
moderate prices and $S$ denotes the regime of price spikes. The MR
model can be defined as
%
%
\begin{eqnarray}
\label{m2} y^A_{M,t}&=&d^A+
\alpha^A y^A_{M,t-1}+\omega^A_{M,t},
\nonumber\\[-8pt]\\[-8pt]
y^A_{S,t}&=&\mu_{S}^A+
\omega^A_{S,t},
\nonumber
\end{eqnarray}
where $A\in\{P,B\}$ refers to the type of aggregation (peakload or
baseload), $\omega^A_{M,t}\sim\mathcal{N}(0,\sigma^2_{M^A})$ and
$\omega^A_{S,t}\sim\mathcal{N}(0,\sigma^2_{S^A})$. The conditional
probabilities of the transitions from one regime to another given the
regime at $t-1$ are captured by the transition matrix
\[
\pmatrix{\mathrm{P} \bigl(R^A_t=M|R^A_{t-1}=M
\bigr) &\mathrm{P} \bigl(R^A_t=M|R^A_{t-1}=S
\bigr)
\cr
\mathrm{P} \bigl(R^A_t=S |R^A_{t-1}=M
\bigr)&\mathrm{P} \bigl(R^A_t=S |R^A_{t-1}=S
\bigr)} =\pmatrix{q&1-p
\cr
1-p&p}
\]
and have to be estimated, too.

\subsubsection*{DSFM} The third model, the DSFM of \citet
{Park2009}, is a functional factor model, which is very similar to our
FFM. Its application to electricity spot prices, as suggested by \citet
{Haerdle2010}, differs from our application, since it models the hourly
spot prices $y_{th}$ based on the classical time series point of view
on electricity spot prices. That is, \citeauthor{Haerdle2010} model
and forecast nonparametric price-hour functions, say, $\chi_t(h)$, and
thereby fail to consider the merit order model. The DSFM can be written as
%
%
\begin{equation}
\label{DSFM} y_{th}=\chi_t(h)+\omega_{th},\qquad
h\in\{1,\ldots,24\},
\end{equation}
with $\chi_t\in L^2[1,24]$ defined as
\[
\chi_t(h)=f^{\mathrm{DSFM}}_0(h)+\sum
_{l=1}^L\beta^{\mathrm{DSFM}}_{tl}f^{\mathrm{DSFM}}_{l}(h),
\]
where $f^{\mathrm{DSFM}}_0(h)$ is a nonparametric mean function,
$f^{\mathrm{DSFM}}_{l}(h)$ are nonparametric functional factors, $\beta
^{\mathrm{DSFM}}_{tl}$ are the univariate scores, and $\omega_{th}$ is a
Gaussian white noise process.

\citeauthor{Park2009} suggest selecting the number of factors $L$ by
the proportion of explained variation. We choose the factor dimension
$\hat{L}=2$, since this factor dimension yields the same proportion of
explained variation as the factor dimension $\hat{K}=2$ for our FFM.

Given the estimates of the time-invariant model components, $\hat
{f}^{\mathrm{DSFM}}_0(h)$, $\hat{f}^{\mathrm{DSFM}}_l(h)$ and $\hat{L}$, forecasting of
the daily price-hour functions $\chi_t(h)$ can be done by forecasting
the estimated univariate time series of scores. As for our FFM, we use
SARIMA models to forecast the univariate time series $(\hat\beta
_{t1}^{\mathrm{DSFM}})$ and $(\hat\beta_{t2}^{\mathrm{DSFM}})$, where the model
selection procedure for the SARIMA models is the same as for our FFM.

\subsubsection*{SFPL} The fourth model, the SFPL model of \citet
{vilar2012forecasting}, is a very recent functional data model, which
was exclusively designed for forecasting electricity spot prices. The
SFPL has the nice property of allowing us to include the values of
electricity demand $u_{th}$ as additional co-variables. \citeauthor
{vilar2012forecasting} use dummy variables for work days and holidays
as additional covariates, which we do not have to do since we consider
only work days. Note that, like the DSFM, the SFPL model uses
price-hour functions $\chi_t(h)$ and therefore does not consider the
merit order model. The definition of the SFPL is given in the following:
%
%
\begin{equation}
\label{SFPL} y_{t+\ell,h}=\alpha u_{th}+m\bigl(
\chi_t(h)\bigr)+\omega_{th},
\end{equation}
where $m\dvtx L^2[1,24]\rightarrow\mathbb{R}$ is a function that maps the
price-hour function $\chi_t$ to a real value and $\omega_{th}$ is a
Gaussian white noise process.

%
%
\begin{figure}[b]

\includegraphics{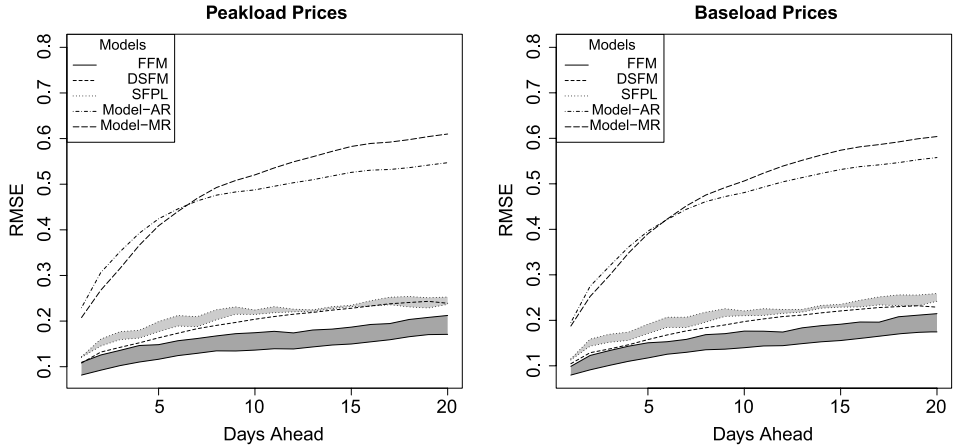}

\caption{Root mean squared errors of the FFM (solid lines) and the
alternative models, DSFM (short-dashed lines), SFPL (dotted lines), AR
(dash-dotted lines) and MR (long-dashed lines) for peakload prices
$y_t^P$ (left panel) and baseload prices $y_t^B$ (right panel). The
gray shaded regions for the FFM and the SFPL model are lower bounded
based on the ideal demand forecast, and upper bounded based on the
persistence forecast.}
\label{RMSE}
\end{figure}
%
%

Forecasting electricity spot prices $y_{th}$ with the SFPL model can be
easily done using the R package \texttt{fda.usc} of \citet{fdausc}.
However, in order to compute the forecasts $\hat{y}_{T_L,h}(\ell
|\mathrm{SFPL})$ of the electricity spot prices $y_{T_L+\ell,h}$, we
also need forecasts of the electricity demand values $u_{T_L+\ell,h}$.
We cope with this problem as suggested above for our FFM by using a
persistence forecast and an ideal forecast.

\subsection{Evaluation of forecast performances}\label{fp}
The two plots of Figure~\ref{RMSE} show the values of the RMSEs for
the $\ell\in\{1,\ldots,20\}$ days ahead forecasts of the peakload
prices (left panel) and the baseload prices (right panel). The two gray
shaded regions in each plot show the possible RMSE values of the FFM
(solid line borders) and the SFPL model (dotted line borders). The
lower bounds of the regions are based on the ideal electricity demand
forecast $\hat{u}^{\mathrm{ideal}}_{T_L,h}(\ell)$. The upper bounds are based
on the persistence forecast of electricity demand $\hat
{u}^{\mathrm{persi}}_{T_L,h}(\ell)$.

The poor performance of the two classical time series models, AR and
MR, in comparison to the three FDA models, FFM, DSFM and SFPL, can be
explained by the different approaches to model the aggregated peakload
and baseload prices. The two classical models try to forecast the
aggregated prices directly, whereas the three FDA models try to
forecast the hourly electricity spot prices; aggregation is done afterward.

The superior performance of our FFM in comparison to the other two FDA
models, DSFM and SFPL, can be explained by the FFMs explicit
consideration of the merit order model. Both models, the DSFM and the
SFPL, work with daily price-hour functions $\chi_t(h)$, which are
based on a rather simple transfer of the classical time series point of
view to a functional data point of view. By contrast, the FFM works
with daily price-demand functions $X_t(u)$, which are based on the
merit order model, the most important model for explaining electricity
spot prices; see our discussion in Section~\ref{i}. Finally, the DSFM
generally performs better than the SFPL model. This might be explained
by the fact that the SFPL model of \citet{vilar2012forecasting} is an
autoregressive model of order one. \citeauthor{vilar2012forecasting}
do not discuss the possibility of extending the order structure of
their SFPL model.

The above study of the RMSEs only gives us insights into the forecast
performances with respect to point forecasts. In order to complement
the forecast comparisons, we also consider interval forecasts. In this
regard, the interval score, proposed by \citet{gneiting2007strictly},
is a very informative statistic. The interval score can be defined as
\begin{eqnarray*}
S^{\mathrm{int}}_\alpha(h,\ell) &=& (\hat{b}_u-
\hat{b}_l)+\frac{2}{\alpha}(\hat{b}_l-y_{T_L+\ell,h})
\mathbb{I}\{y_{T_L+\ell,h}<\hat{b}_l\}
\\
&&{}+\frac{2}{\alpha}(y_{T_L+\ell,h}-\hat{b}_u)\mathbb{I}
\{y_{T_L+\ell,h}>\hat{b}_u\},
\end{eqnarray*}
where $\hat{b}_u=\hat{b}_{u,T_L,h}(\ell)$ and $\hat{b}_l=\hat
{b}_{l,T_L,h}(\ell)$ are the lower and upper bounds of the $(1-\alpha
)\%$ forecast interval for the electricity spot price $y_{T_L+\ell,h}$.
The interval score punishes a broad prediction interval $(\hat
{b}_u-\hat{b}_l)$ and adds an additional punishment if the actual
observation $y_{T_L+\ell,h}$ is not within the prediction interval. In
general, a~lower interval score is a better one.

Unfortunately, we cannot compute the interval scores for all five
models. For example, \citet{vilar2012forecasting} do not propose any
prediction intervals for the SFPL model. Furthermore, while it is easy
to compute forecast intervals of the FFM and the DSFM for hourly sport
prices, it is not trivial to compute them for the aggregated (peakload
and baseload) prices.

Therefore, we focus on the hourly forecasts of electricity spot prices
of the FFM and DSFM models. For both models, the 95\% forecast
intervals can be computed on the basis of the 95\% forecast intervals
of the SARIMA forecasts given the estimated factors.

\begin{figure}

\includegraphics{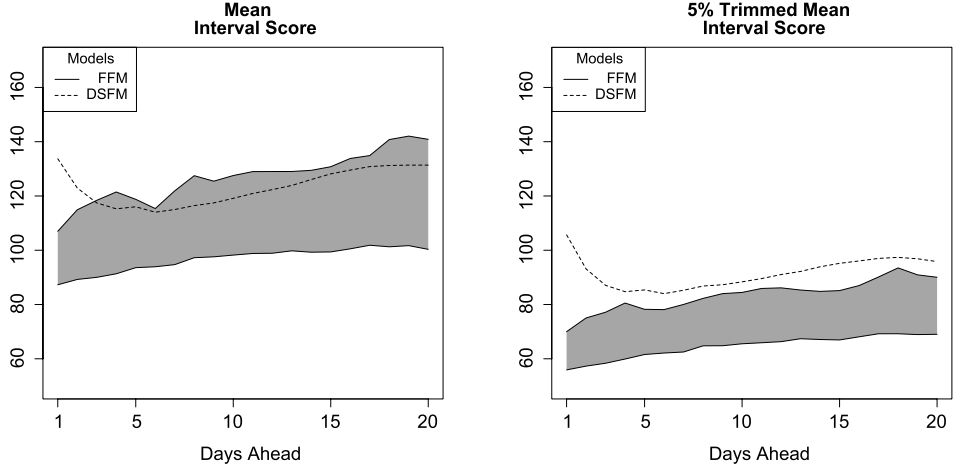}

\caption{Mean and trimmed mean values of the interval scores
$S^{\mathrm{int}}_\alpha(h,\ell)$, pooled for all hours $h\in\{1,\ldots,24\}
$. A low interval score stands for precise predictions with narrow
prediction intervals. The dashed line corresponds to the interval
scores of the DSFM of Park et~al. (\citeyear{Park2009}). The gray shaded regions for the
FFM are lower bounded based on the ideal demand forecast, and upper
bounded based on the persistence forecast.}
\label{IntervalScore}
\end{figure}

Due to the enlargement of the learning sample after each $\ell$ days
ahead forecast by one day and due to pooling all hours
$h\in\{1,\ldots,24\}$, we have for each $\ell$ days ahead forecast
$24\cdot(197-\ell )$ interval scores $S^{\mathrm{int}}_\alpha(h,\ell)$ in order
to compare the $\ell$ days ahead forecast performances of our FFM and
the DSFM. In Figure~\ref{IntervalScore} we present the (trimmed) mean
values of these pooled interval scores $S^{\mathrm{int}}_\alpha(h,\ell)$ with
$\alpha =0.05$ for each $\ell\in\{1,\ldots,20\}$. The $5\%$ trimmed mean
values are used, since for both models there are some extreme values of
the interval score (due to the outliers in the forecast sample), which
distort the mean values.

%


Figure~\ref{IntervalScore} clearly confirms the good forecast
performance of the FFM. Besides some technical issues, the main
conceptual difference between the DSFM and our FFM is that the DSFM
works with daily price-\emph{hour} functions $\chi_t(h)$, whereas our
FFM works with daily price-\emph{demand} functions $X_t(u)$, which are
suggested by the merit order model. This demonstrates that the
consideration of the merit order model yields better point forecasts as
well as better interval forecasts.

\section{Conclusion}
In this paper we suggest interpreting hourly electricity spot prices as
noisy discretization points of smooth price-demand functions. This
functional data perspective on electricity spot prices is motivated as
well as theoretically underpinned by the merit order model---the most
important pricing model for electricity spot prices.

We propose a functional factor model in order to model and forecast the
nonstationary time series of price-demand functions and discuss a
two-step estimation procedure. In the first step we estimate the single
price-demand functions from the noisy discretization points. In the
second step we robustly estimate from these a finite set of common
basis functions. The careful consideration of the merit order model
yields a very parsimonious functional factor model with only two common
basis functions, which together explain over $99\%$ of the total sample
variation of the price-demand functions.

Our approach allows us to separate the total variations of electricity
spot prices into one part caused by the variations of the merit order
curves (mainly variations of input-costs) and another part caused by
the variations of electricity demand. The first part is modeled by our
FFM and the second part can be modeled by specialized methods proposed
in the literature. We decided to keep the model parsimonious;
nevertheless, it is easily possible to include the input cost for
resources (coal, gas, etc.) into our FFM. Researchers are invited to
extend the FFM for these co-variables.

The presentation of our functional factor model is concluded by a real
data application and a forecast study which compares our FFM with four
alternative time series models that have been proposed in the
electricity literature. The real data application demonstrates the use
of the functional factor model and a possible interpretation of the
unobserved common basis functions. The forecast study clearly confirms
the power of our functional factor model and the use of price-demand
functions as underlying structures of electricity spot prices in general.

\section*{Acknowledgments}

I want to thank Alois Kneip (University of Bonn) and Pascal Sarda
(Universit\'e Paul Sabatier, Toulouse) for stimulating discussions.
Section~\ref{f} profited especially from comments of the referees.

\begin{supplement}
\stitle{R-codes and data set}
\slink[doi]{10.1214/13-AOAS652SUPP} 
\sdatatype{.zip}
\sfilename{aoas652\_supp.zip}
\sdescription{In this supplement we provide a zip file containing the
R-Codes and the data set used to model and forecast electricity spot
prices by the functional factor model as described in this paper.}
\end{supplement}


\printaddresses

\end{document}